\documentclass[11pt]{article}
\usepackage{moriond,epsfig}

\def\Journal#1#2#3#4{{#1} {\bf #2}, #3 (#4)}

\def\PRL{\em Phys. Rev. Lett.}
\def\PRD{{\em Phys. Rev.} D}

\def\be{\begin{equation}}
\def\ee{\end{equation}}
\def\bea{\begin{eqnarray}}
\def\eea{\end{eqnarray}}

\begin{document}
\vspace*{4cm}
\title{TOP QUARK AND ELECTROWEAK RESULTS FROM DZERO}

\author{H.~T.~DIEHL}

\address{Fermi National Accelerator Laboratory, Batavia, IL, 60510 USA}

\maketitle

\abstracts{Reported are new top quark and electroweak results 
from the D\O \ experiment, which is accumulating luminosity from $p\bar{p}$ 
collisions at center-of-mass energy 1.96 TeV (Run~II).}

%\section{Introduction}
%I discuss new top quark and electroweak results 
%from the D\O \ experiment, which is at present accumulating luminosity 
%from $p\bar{p}$ collisions at center-of-mass energy 1.96 TeV (Run~II). 
%The new results include the $W$ and $Z$ boson cross sections in Run~II, 
%a search for heavy neutral gauge boson that couples to electrons, 
%and the $t\bar{t}$ production cross section in six decay channels in
%Run~II. Lastly, I describe a new preliminary result on the top quark
%mass, as determined from the Run~I data.

\section{$W$ and $Z$ Production}
We study $W$ and $Z$ boson production at the Tevatron to extract a 
variety of physics. For instance, comparison of the branching fractions
$B(W\rightarrow \ell \nu)$, where $\ell=$e, $\mu$, or $\tau$, tests the
universality of the leptonic couplings to the weak current~\cite{D0tau}. 
The angular distribution of leptons from $W$ boson decay provides 
constraints on parton distribution functions that describe the structure of 
the proton~\cite{cdfwa} and allow us to understand beyond-tree-level 
QCD corrections to production models~\cite{D0a2}. 
Equally importantly at this early stage in Run~II, 
these relatively common processes allow us to benchmark the performance 
of our detector, trigger, and reconstruction algorithms. 

\begin{figure}
\psfig{figure=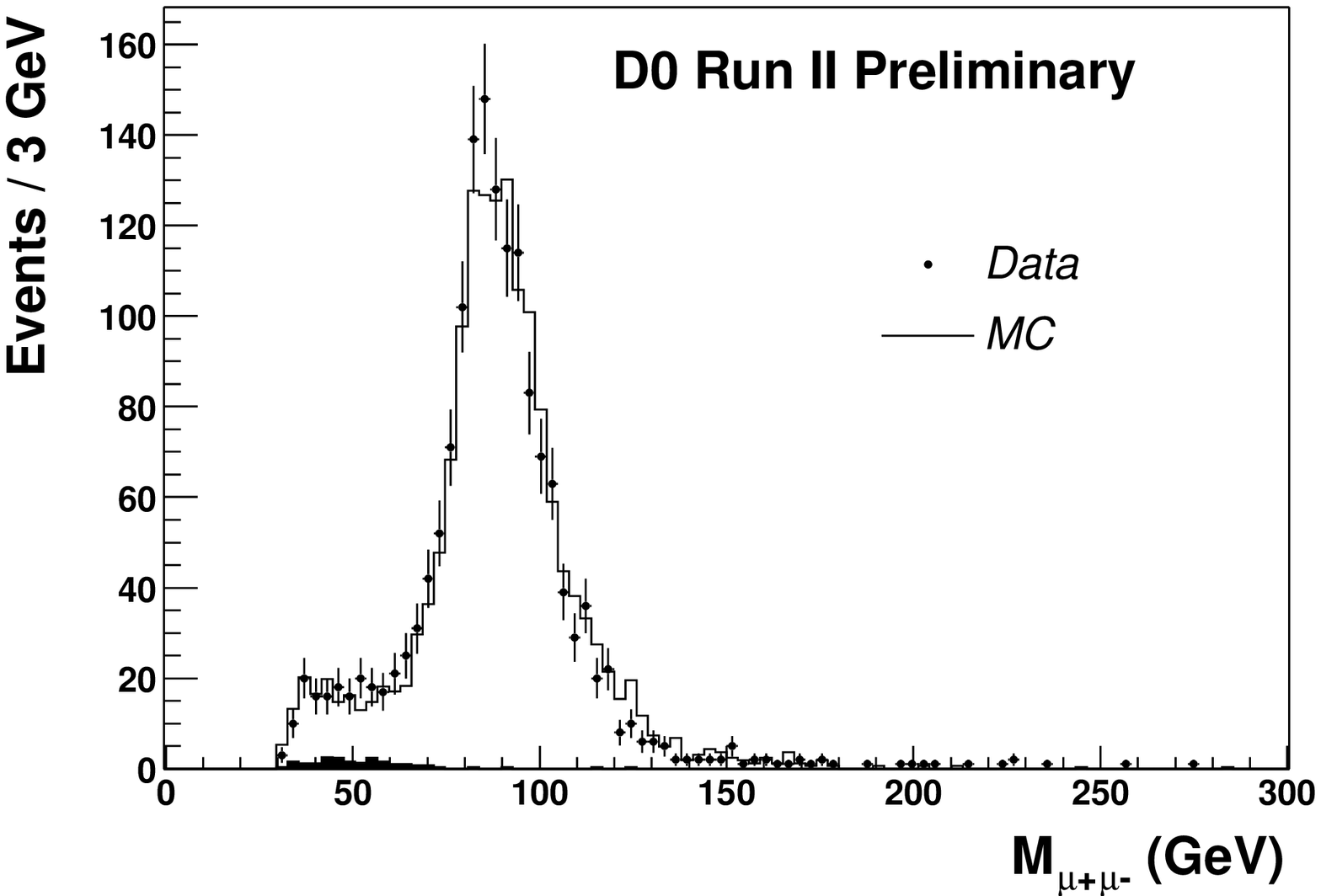,height=2.2in}
\psfig{figure=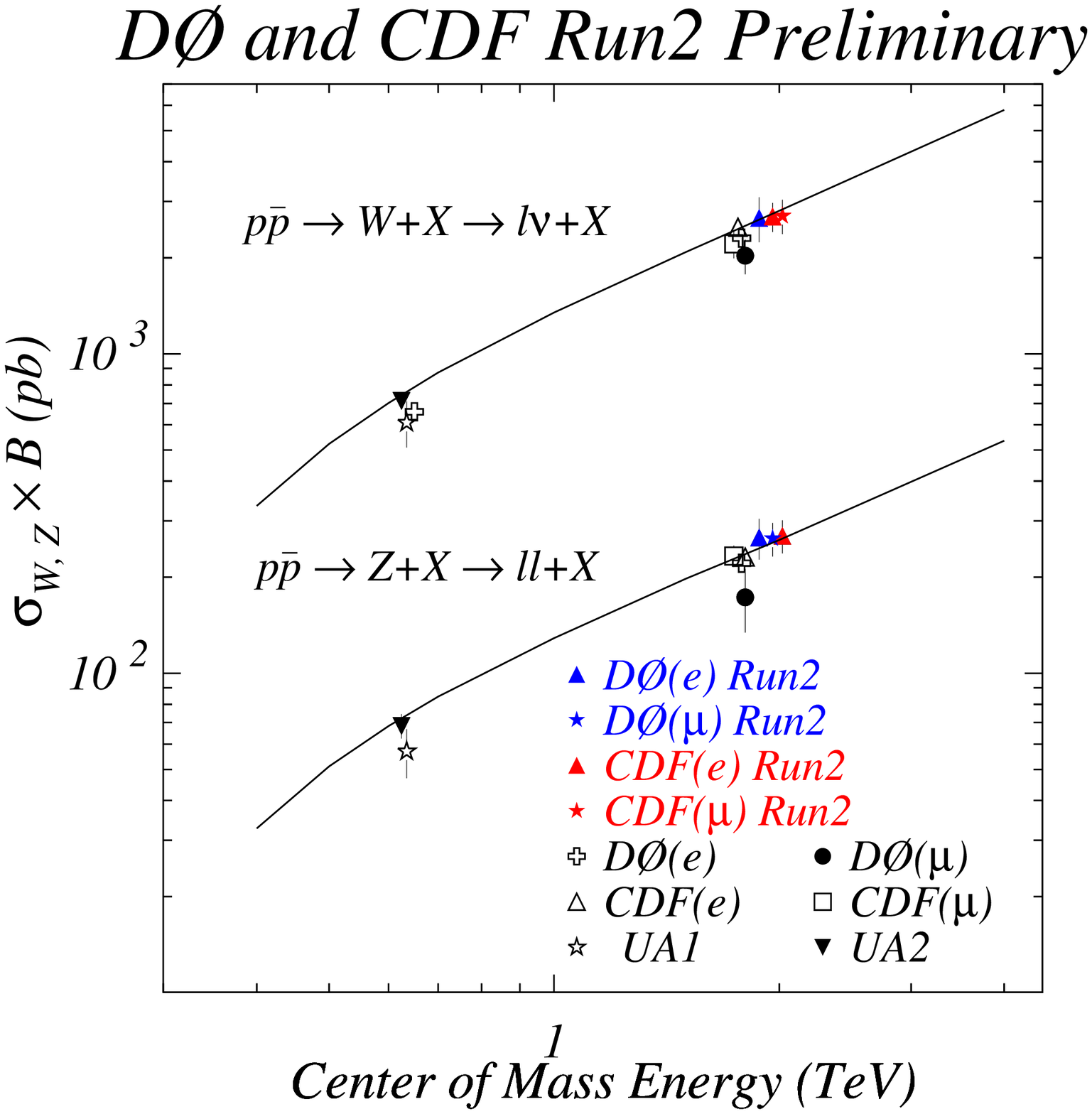,height=2.5in}
\caption{The figure on the left is the dimuon invariant mass plot 
in 32 pb$^{-1}$ (data is points) showing the $Z\rightarrow \mu\mu$ peak 
and Drell-Yan tail. The small shaded region is the expected background. 
The histogram is MC.
On the right is $\sigma Br$ for $W$ and $Z$ bosons in the 
e and $\mu$ decay modes as a function of center-of-mass energy 
at $p\bar{p}$ colliders.}
\label{fig:zmuons}
\end{figure}

Using $32$ pb$^{-1}$ of $p\bar{p}$ collisions at  $\sqrt{s}=1.96$ TeV, we 
measured the cross section times branching ratio 
$\sigma(Z+X)Br(Z\rightarrow \mu \mu)$. The selection 
criteria required at least two oppositely-charged muons with $pT\ge 15$ GeV/c
within $|\eta| \le 1.8$ and $(\Delta R)^2=(\Delta \phi_{\mu\mu})^2 +
(\Delta \eta_{\mu\mu})^2 \ge 4.0$. 
At least one muon was required to be isolated in the calorimeter and the
central tracker.
Figure~\ref{fig:zmuons}a shows the dimuon invariant mass distribution
for the resulting 1585 muon pairs (data points) and the 
$~1.5\pm 1.0\%$ background from $b\bar{b}$ and 
$Z\rightarrow \tau \tau$ events (shaded). The histogram is the Monte Carlo.
Because there was no explicit mass selection, the Drell-Yan contribution,
determined using Pythia, is $12\pm 1\%$. It was $4\%$ in the region 
$75 \le M(\mu\mu) \le 105$ GeV/c$^2$.
The result is $\sigma(Z+X)Br(Z\rightarrow \mu \mu)=264\pm7(stat.)\pm 
17(sys.)\pm26(lum'y)$ pb. The largest uncertainty, 26 pb, comes from 
uncertainty in the delivered luminosity.

Figure~\ref{fig:zmuons}b shows the cross section times branching ratio 
for $W$ and $Z$ bosons in the electron and muon channels in $p\bar{p}$ 
collisions as a function of center-of-mass energy including the new result
(described above), preliminary results~\cite{cdfr2}$^-$\cite{mk} from Run~II,
and previously published results~\cite{ua-d0}. Comparing the cross sections
times branching ratios measured by D\O \ at 1.8 TeV and 1.96 TeV, we find 
$\sigma_{W\rightarrow e\nu}(1960 \; \rm{GeV})/\sigma_{W\rightarrow e\nu}(1800
\; \rm{GeV})=1.15\pm0.19$, $\sigma_{Z\rightarrow ee}(1960 \; \rm{GeV})/
\sigma_{Z\rightarrow ee}(1800\; \rm{GeV})=1.20\pm0.19$, and 
$\sigma_{Z\rightarrow \mu\mu}(1960 \; \rm{GeV})/\sigma_{Z\rightarrow \mu\mu}
(1800 \; \rm{GeV})=1.48\pm0.32$. In the first two cases the main uncertainty
is the Run~II luminosity. In the case of $Z$ to muons, the main uncertainty
is in the Run~I result, where we struggled with the muon acceptance. In 
fact, the $Z$ boson peak in Fig.~\ref{fig:zmuons}a indicates the success
of the D\O \ Upgrade for Run~II.

\section{Search For $Z'\rightarrow$ Di-electrons}
\begin{figure}
%\rule{5cm}{0.2mm}\hfill\rule{5cm}{0.2mm}
%\vskip 2.5cm
%\rule{5cm}{0.2mm}\hfill\rule{5cm}{0.2mm}
\psfig{figure=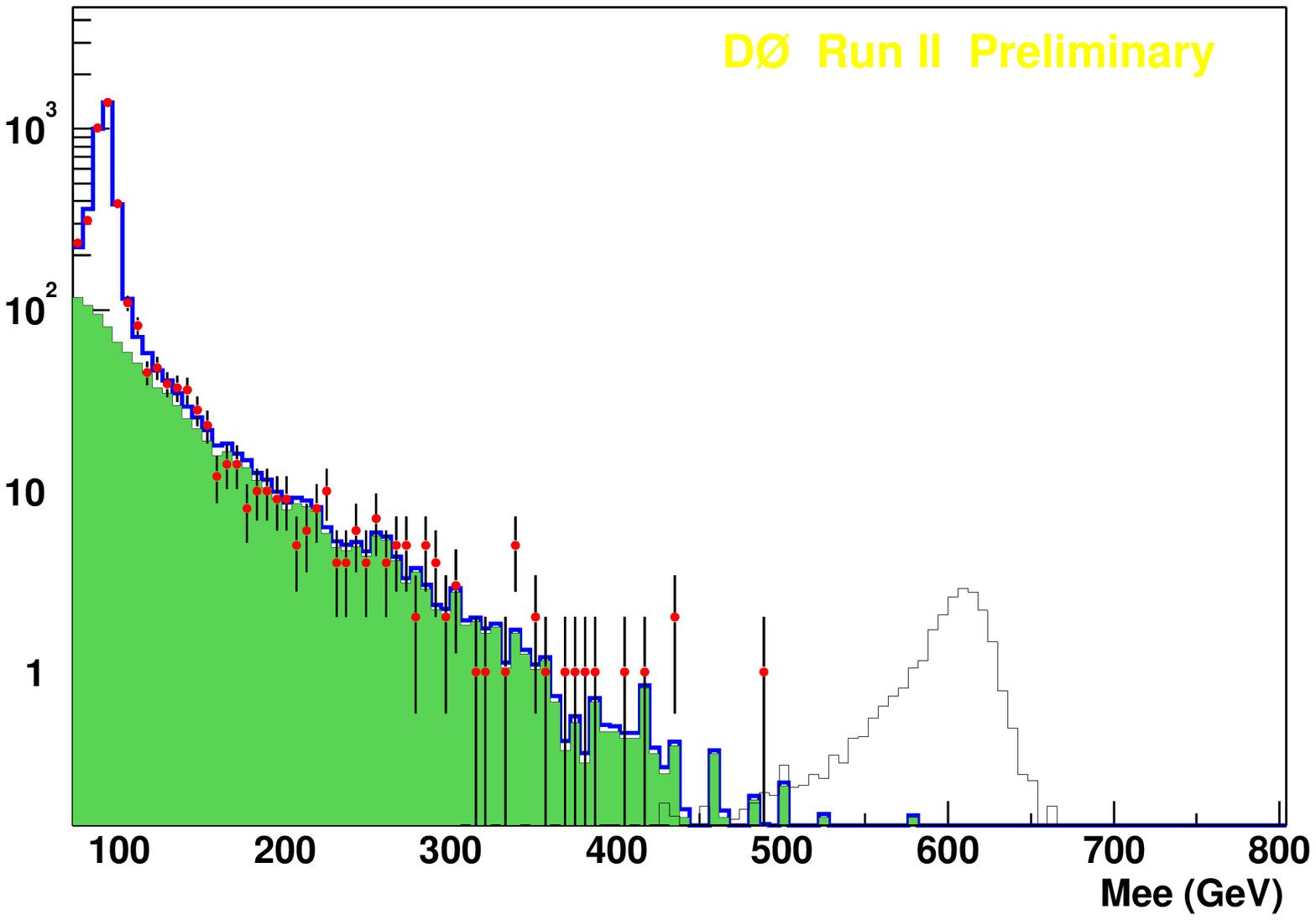,height=2.2in}
\psfig{figure=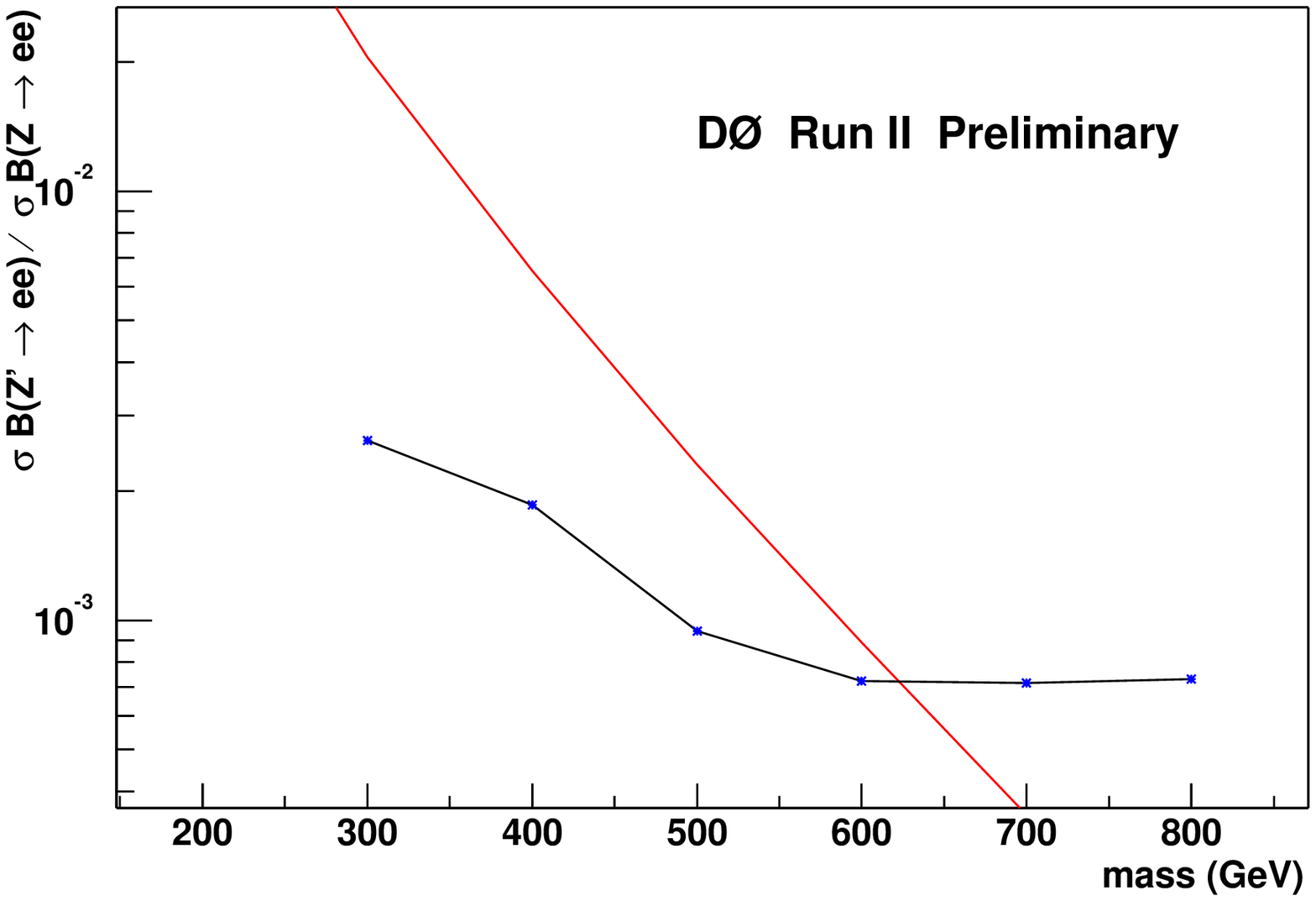,height=2.2in}
\caption{On the left is the di-EM invariant mass plot. The (red) points are
the data. The histogram is the $Z$ boson and Drell-Yan electrons from Monte
Carlo. The (green) shaded region is hadronic background. Also displayed is 
a MC $Z'$ with mass 600 GeV/c$^2$ enhanced by 10 times. On the 
right is the 95\% C.L. cross section limits (points) and theory
prediction (red) 
for production as a function of $Z'$ mass. $M(Z')\ge 620$ GeV/c$^2$.}
\label{fig:zprime}
\end{figure}

Using 50 pb$^{-1}$ we searched for non-SM particles that decay to 
lepton-antilepton pairs. We assume the couplings of the leptons to
the putative particle are the same as their couplings to the $Z$ boson and
we refer to it as a $Z'$.  We triggered on events with EM clusters within
the region $|\eta|\le 0.8$.  The remaining selection criteria are designed
not to remove very high $E_T$ electrons. We simply require events with 
two or more isolated EM showers with $E_T \ge 25$ GeV in the fiducial regions 
$|\eta|\le 1.1$ or $1.5\le |\eta|\le 2.5$.  Figure~\ref{fig:zprime}a shows 
the di-EM invariant mass from 80 to 800 GeV/c$^2$.  After
subtracting the multijet background we observe 2817 $Z\rightarrow ee$ and 
Drell-Yan events.

No excess of events was observed at any mass. We set limits on the ratio 
of cross section for $Z'$ production compared to the $Z$ boson production
(so that many systematic uncertainties are removed) as a function of the 
putative $Z'$ mass. Figure~\ref{fig:zprime}b 
shows the 95\% C.L. limits and the 
(default) Pythia prediction as a function of mass. The two lines cross
at 620 GeV, the 95\% C.L. mass limit.

\section{Top Quark Cross Section at $\sqrt{s}=1.96$  TeV}
We have measured the $t\bar{t}$ cross section at $\sqrt{s}=1.96$ TeV.
Top (antitop) quarks decay to a $W^{+(-)}$ boson and a (anti)$b$-quark 
and the different decay channels are ``named'' by the $W$ boson decay 
modes and whether or not the $b$-quarks are tagged with either a secondary 
vertex or a muon.  We combined the results from six decay channels:
``electron plus jets'', ``electron plus jets with a muon tag'',
``muon plus jets'' (with and without a muon tag), and the $\mu\mu$ and 
e$\mu$ ``dilepton'' modes. 

In the ``leptons plus jets'' analysis we preselected a sample rich in $W$
boson decays to an electron or muon (and neutrino) by requiring an isolated
lepton with $pT \ge 20$ GeV/c and at least 20 GeV of missing transverse 
energy (MET). We veto on events with a non-isolated muon, preserving them 
for the tagged-sample analysis. We bin them according to the number of jets 
with $E_T \ge 15$ GeV and evaluate the hadronic background 
in each bin. The data samples correspond to 50 (40) pb$^{-1}$ of 
collisions in the electron (muon) channels. 
%Figure~\ref{fig:berends} 
%shows the number of events, after removing the hadronic background, in the 
%electron plus jets and muon plus jets channels, as well as the 
%line fit to the number of $W$ plus one or more, two or more, and three or more
%jets events.  
We estimate that the number of hadronic background and 
$W$+4 or more jets events in the 
electron + jets (muon + jets) channel is $12.5$ $(11.9)$ and $11.9$ (24.2), 
respectively. We observe 22 (38)
events.  To further distinguish 
the backgrounds due to hadrons and $W$+jets from the top signal we apply
topological selection criteria including, variously, the $E_T$ of the 
highest $E_T$ jet, the $E_T$ and pseudorapidity of the reconstructed 
$W$ (dijets) boson, the total hadronic $E_T$ (called $H_T$), and the 
aplanarity. In the electron plus jet analysis the bottom line is
$2.7\pm0.6$ background events expected, 1.8 $t\bar{t}$ events (assuming
$\sigma(t\bar{t})$ is 7 pb), and 4 candidates observed. In the muon plus
jets channel, the corresponding numbers are $2.7\pm1.1$ background,
2.4 signal, and 4 candidates observed.

In the ``lepton plus jets with muon tag'' samples, we start with the 
preselected events described above that were vetoed because of the 
non-isolated muon. We further require the events have at least 3 jets with 
$E_T \ge 20$ GeV within $|\eta| \le 2.0$, that there be at least 
110 GeV of $H_T$, and that the events be aplanar. The bottom line
is that in the electron plus jets with muon tag analysis, we expect 
$0.2\pm 0.1$ background events, 0.5 $t\bar{t}$ events (again assuming
$\sigma(t\bar{t})$ is 7 pb), and we observe 2 candidates. 
In the muon plus jets with muon tag analysis, we expect 
$0.6\pm 0.3$ background events, 0.4 $t\bar{t}$ events, 
and we observe no candidates. 

%\begin{figure}
%%\psfig{figure=topxsec.eps,height=2.2in}
%\psfig{figure=BerendsAll.eps,height=2.2in}
%\caption{The number of $W$ plus N (or more) jets in the muon plus jets
%(left) and electron plus jets (right) analyses. The projection of the line
%fitted through N = 1 to 3 provides the $W$+jets background in the 4th bin.}
%%\label{fig:topxsec}
%\label{fig:berends}
%\end{figure}

The data in the $\mu\mu$ and e$\mu$ ``dilepton'' decay channels
corresponds to 43 and 33 pb$^{-1}$, respectively. 
In the dimuon channel we require at least two isolated muons with 
$p_T\ge 15$ GeV/c, two or more jets with $E_T\ge 20$ GeV, MET $\ge 30$
GeV (except around the $Z$ boson mass where we require at least 40 GeV),
and that the $H_T$ be at least 100 GeV. The remaining background is
a combination of $Z$ boson, Drell-Yan dimuon, and $b$-quark decays and 
is expected to total $0.6\pm0.3$ events. SM top, with the usual assumption
for the cross section, is expected to produce $0.30\pm0.04$ events. We
observe 2 candidates. In the e$\mu$ channel we require an isolated electron
and muon with $p_T\ge 15$ GeV/c, that the muon-corrected MET be at least 
10 GeV, the not-muon-corrected MET be at least 20 GeV, two or more jets 
with $E_T\ge 20$ GeV, and $H_T \ge 120$ GeV. The expected background of
$Z\rightarrow \tau\tau$ and hadronic events is expected to result in
$0.07\pm0.01$ candidates. SM top is expected to yield $0.5\pm0.1$ 
candidates.  We observe one candidate. 

When we combine the results in all six channels, a $3\sigma$ excess exists 
in the number of events observed compared to background expected. 
Channel-to-channel the observed distribution is consistent with SM 
top at the 35\% C.L., consistent with interpretation of the signal as SM
$t\bar{t}$ production. The cross section is $8.4^{+4.5}_{-3.7}(stat.)
^{+5.3}_{-3.5}(sys.)\pm0.8 (lum'y)$ pb. This represents a 47\% increase 
over the cross section at 1.8 TeV~\cite{d01atop}. 

\section{Improved Top Quark Mass Measurement}
We report a new measurement of the top quark mass extracted from the
125 pb$^{-1}$ sample accumulated during Run~I. 
This measurement is an update over that which we have published 
previously~\cite{topmass} from the leptons plus jets data, where the final 
data sample consisted of 91 events with an isolated lepton and 4 or more jets. 

Some additional event selection was applied in the new analysis. The 
candidates were required to contain exactly 4 jets for comparison,
on an event-by-event basis, with a leading-order matrix element 
calculation for the
production and decay process. That reduced the sample to 71 events.
The $W$+jets background probability was determined from VECBOS matrix 
elements. Selecting on background probability 
reduced the sample to 22 events with signal efficiency of 70\%.

A likelihood variable~\cite{estrada} 
involving all of the measured features of the 
event was formed from the probability the event was $W$+jets background
or top signal of a given mass.  The probability for top signal was 
determined from the leading-order matrix element calculation including all 
12 jet-identification permutations, all possible values of the neutrino 
momentum and the detector response as measured from the data. The 
likelihood is plotted as a function of top quark mass. The maximum value 
of the likelihood occurred with 12 of the 22 events determined to 
be signal, 10 to be background, and a top mass 
$M_{top}=179.9\pm3.6(stat.)\pm6.0(sys.)$ GeV/c$^2$. 
The systematic uncertainty is dominated by 
contributions from uncertainty in the jet energy scale (5.6 GeV/c$^2$).

This technique provides great promise for Run~II because of it's statistical
power (the statistical uncertainty is reduced by nearly a factor of
1.6 from the Run~I result) and because the main systematic is constrained 
by the $W$ boson mass as determined from the dijet invariant mass 
combination in the signal. 

%\begin{figure}
%\begin{center}
%\setlength{\unitlength}{1mm}
%\begin{picture}(80,80)(0,0)
%\put(0,0){\includegraphics[width=.6\textwidth]{discr.eps}}
%\put(10,35){\includegraphics[width=.5\textwidth]{discr_old.eps}}
%\end{picture}
%\end{center}
%\caption[]{Discriminator $D=P_{signal}/(P_{signal}+P_{backg.})$ for the
%new analysis, and (inset) for the two versions of the
%analysis included in the previous D\O\ publication \cite{massPRD5}.}
%\label{fig:discr}
%\end{figure}

\section{Summary}
We have reported $W$ boson, $Z$ boson, and $t\bar{t}$ cross section 
results.  They demonstrate the upgraded D\O \ 
detector is performing nicely in Run~II. We have presented an 
update of D\O\'s Run~I top mass measurement in the lepton plus
jets channel. The technique provides great promise for Run~II.

\section*{Acknowledgments}
I am grateful for help from E.~Nurse, G.~Steinbrueck, M.~Verzocchi, 
M.~C.~Gao, E.~Barberis, C.~Gerber, M.~Klute, M.~Kado, and J.~Estrada 
in preparation of this talk.

%\section*{Appendix}

\end{document}